\documentclass{aastex631}
\usepackage{amsmath}
\usepackage{float}

\shorttitle{Loop projection effects}
\shortauthors{Uritsky et al.}

\newcommand{\vmu}[1]{{{#1}}}
\newcommand{\jak}[1]{{{#1}}}

\begin{document}

\title{Are coronal loops projection effects?}

\correspondingauthor{Vadim Uritsky}
\email{vadim.uritsky@nasa.gov}

\author[0000-0002-5871-6605]{Vadim M. Uritsky}
\affiliation{Catholic University of America\\
Washington, DC 20064, USA}
\affiliation{NASA Goddard Space Flight Center\\
Greenbelt, MD 20771, USA}

\author[0000-0003-2255-0305]{James A. Klimchuk}
\affiliation{NASA Goddard Space Flight Center\\
Greenbelt, MD 20771, USA}

\begin{abstract}

We report results of an in-depth numerical investigation of three-dimensional projection effects which could influence the observed loop-like structures in an optically thin solar corona. Several archetypal emitting geometries are tested, including collections of luminous  structures with circular cross-sections of fixed and random size, light-emitting structures with highly anisotropic cross-sections, as well as two-dimensional stochastic current density structures generated by fully-developed magnetohydrodynamic (MHD) turbulence. A comprehensive set of statistical signatures is used to compare the line of sight -integrated emission signals predicted by the constructed numerical models with the loop profiles observed by the extreme ultraviolet telescope onboard the flight 2.1 of the High-Resolution Coronal Imager (Hi-C). \vmu{The results suggest that typical cross-sectional emission envelopes of the Hi-C loops are unlikely to  have high eccentricity, and that the observed loops cannot be attributed to randomly oriented quasi-two dimensional emitting structures, some of which would produce anomalously strong optical signatures due to an accidental line-of-sight alignment, as expected in the coronal veil scenario} \citep{malanushenko2022}. The possibility of apparent loop-like projections of very small (close to the resolution limit) or very large (comparable with the size of an active region) light-emitting  sheets remains open, but the intermediate range of scales commonly associated with observed loop systems is most likely filled with true quasi-one dimensional (roughly axisymmetric) structures embedded into the three-dimensional coronal volume.

\end{abstract}


\section{Introduction} \label{sec:intro}
\label{sec:introduction}

Any successful theory of coronal heating must explain how the heating is distributed in space and time. Because the highly conducting coronal plasma is constrained to flow along the magnetic field, and because thermal conduction is highly anisotropic – being strong along the field and weak across it – magnetic flux “tubes” in the coronal volume tend to have a uniform temperature and density, which can be much different from the temperature and density of adjacent tubes. It is variations in the heating that give rise to these thermodynamic differences. Thus, elongated intensity features in coronal images, i.e., coronal loops, are a direct consequence of cross-field spatial variations in heating. By studying the cross sections of the emitting structures responsible for the loops, we can gain valuable information about coronal heating and establish crucial observational tests for heating theories. The challenge is that we do not observe the 3D shapes of the emitting structures. Rather, we observe their projection onto the 2D plane of an image. Additional complication comes from line-of-sight overlap. The corona is optically thin, so the intensity observed in an image pixel is an integration of all the emission along the line of sight. 

Traditionally, and for no compelling reason, the loops seen in coronal images have been assumed to correspond to flux tubes with circular cross sections. This is the shape that comes to mind with the term “tube.” Attempts to verify or rule out a circular shape using observations from orthogonal vantage points have not been definitive due to the difficulty in identifying the same loop in the two images \citep{west2014, kuchera2019,mccarthy2021}. 
Potential field models suggest that coronal flux tubes do not maintain the same transverse geometry through their length, challenging the circular cross section assumption \citep{malanushenko2013}.

\citet{klimchuk2020} used a different scheme to address the circularity of the cross section. They noted that there is abundant evidence for twist in the corona and pointed out that a twisted flux tube with a noncircular cross section would exhibit an anti-correlation between the intensity and width along the loop in an image. For some portions of the loop, the line of sight would be aligned with the long dimension of the cross section and the loop would appear bright and thin, while for other portions the line of sight would be aligned with the short dimension and the loop would appear faint and fat. \citet{klimchuk2020} studied 20 long loop segments observed in the first flight of the Hi-C rocket experiment \citep{winebarger2014} and found no evidence for an anti-correlation. They concluded that the cross sections must be roughly circular. The emission can be nonuniform within the cross section and the envelope can have an irregular shape, but it cannot have a large aspect ratio. \citet{williams2021} performed a similar analysis on data from the third Hi-C flight and obtained a similar result. Note that simple shapes, such as ellipses, produce a linear anti-correlation between intensity and width, while more complex shapes can produce a nonlinear anti-correlation. In all cases there is an anti-correlation if the aspect ratio is large.

MHD simulations make different predictions about the shapes of loop cross sections. \citet{knizhnik2018} simulated what would correspond to a small portion of an active region and emphasized the multi-strand nature of the corona. \jak{The magnetic field is highly inhomogeneous in the photosphere and tends to be concentrating in small patches of kilogauss strength. Emanating from each patch is what can be considered a quasi-distinct magnetic flux tube, which we refer to as a coronal strand. Observations show that a single active region contains in excess of 100,000 of these strands} \citep{klimchuk2015}. The strands become twisted and tangled by chaotic photospheric convection, and therefore the boundaries between them are current sheets. Reconnection at these current sheets gives rise to nanoflare heating events. It is impossible at present to simulate all the current sheets in an active region, so Knizhnik et al. modeled an idealized subset of an active region in order to include the fundamental multi-strand, multi-current sheet nature of the field. The pattern of emission in their simulation suggests loops with approximately circular cross sections \citep{klimchuk2023}. Each loop is a bundle of strands where the nanoflares exhibit a collective behavior. They are the sites of nanoflare “storms.” Although the imposed driver flows in the simulation have a circular nature, they do not have a simple mapping to the circular emission patches associated with loop cross sections. The driver cells have a smaller scale, and multiple different cells are linked to each patch. Furthermore, recent simulations not yet published indicate that translational driver flows also produce circular cross sections. 

Another recent simulation considers an entire active region \citep{malanushenko2022}. Although it does not include the multitude of small-scale strands and current sheets of a real active region, it incorporates the large-scale stresses and currents that are missing from the \citet{knizhnik2018} simulation. \citet{malanushenko2022} find that many of the loops in the synthetic images based on their model correspond not to confined emission structures, i.e., “tubes,” but rather to large warped 2D structures within the 3D volume. They call these large structures “veils.” Bright loops appear in places where the observer is looking along a veil and the line-of-sight depth is large. Elsewhere the observer is looking through the thin veil and the intensity is faint. The effect is very clear.  However, the heating that produces the veils in the simulation has yet to be investigated. It is likely due to relatively passive ohmic and viscous dissipation of large-scale current and velocity structures, which can give significant heating because of the  much-smaller-than-solar Reynolds numbers of the simulation. Whether this is a good proxy for the explosive energy release that occurs at small-scale current sheets in the real, multi-stranded corona has yet to be established. 

In this paper we report on a new study to determine observationally the cross-sectional properties of emitting structures in the corona. By analyzing intensity profiles from cuts across the magnetic field in both artificial and real coronal images, and by applying a variety of statistical techniques, we are able to infer the shapes and size distributions of the cross sections. Our results provide valuable tests for MHD simulations and place important constraints on theories of coronal heating.

This paper is organized as follows. Section 2 describes the simulation techniques that we developed for reproducing some common projection effects influencing loop observations. Section 3 presents a toolkit of data analysis methods enabling an empirical classification of the  simulated loops systems based on their transverse profiles. In Section 4, this data analysis methodology is applied to the systems of loops observed by the extreme ultraviolet (EUV) telescope onboard the Hi-C 2.1 sounding rocket experiment. Section 5 compares the numerically simulated loops system with the observed loops, which allows us to infer the cross-sectional shape of the Hi-C loops and to evaluate the role of projection effects in the optically thin corona.

\section{Simulation techniques }
\label{sec:models}


\subsection{The Stochastic Pulse Superposition framework}
\label{sec:sps}    

\begin{figure}
\begin{center}
\includegraphics[width=8.0 cm]{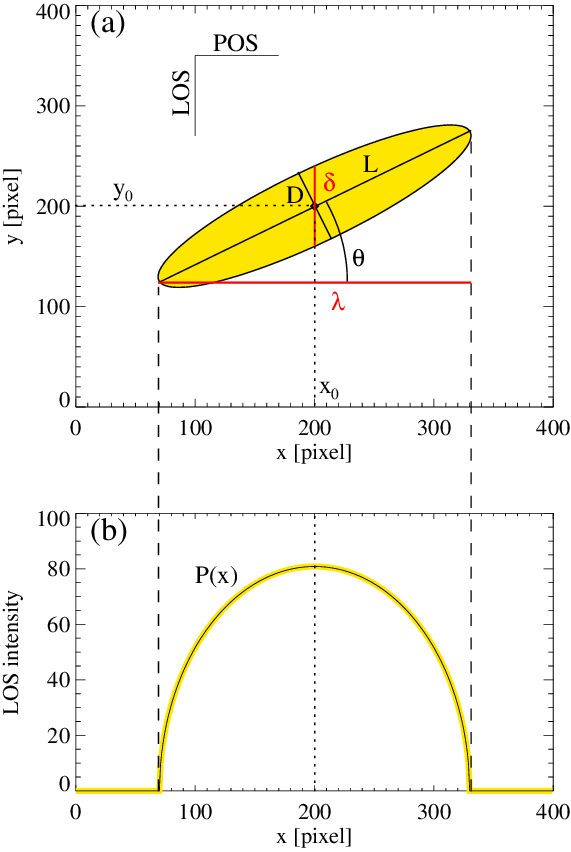}
\caption{\label{fig_pulse} The building block of the SPS framework -- an elliptical light-emitting structure representing a cross section of an anisotropic loop strand (a) producing a LOS-integrated emission pulse $P(x)$ (b). $L$ and $D$ are correspondingly the major and minor axes of the structure, $\lambda$ is its projected size, $\delta$ is the maximum LOS depth of the structure, $x_0$, $y_0$ and $\theta$ are respectively the coordinates and the orientation angle of the structure. The PSP models consist of a large number pulses such as the one illustrated here, generated and positioned randomly according to a set of chosen probabilistic rules.}
\end{center}
\end{figure}

The main element of the Stochastic Pulse Superposition (SPS) simulation framework is a randomly positioned, randomly oriented  elliptical structure representing the cross-sectional shape of a single luminous loop or a loop strand, as shown schematically on Figure \ref{fig_pulse}(a). The major axis of the ellipse $L$ is interpreted as the width of the structure and the minor axis $D$ defines its thickness. The relationship between $L$ and $D$ controls a scale-dependent aspect ratio of the structures as discussed below. The structures are embedded into an $(x,y)$ coordinate plane assumed to be perpendicular to the overall magnetic field direction of the simulated loop system. (Individual magnetic strands may deviate in orientation.) The $x$ coordinate is parallel to the plane of sky (POS) and defines a virtual slit across the loop, while the $y$ axis  is aligned with the line of sight (LOS) of the virtual observer. The central coordinates $x_0$, $y_0$ and the orientation angle $\theta$ of the structures were sampled from three independent uniform probability distributions, with the supports $[0, N_x-1]$, $[0, N_y-1]$ and  $[- \pi  , \pi]$, correspondingly.

The projected POS size $\lambda$ and the maximum LOS depth $\delta$ of the structure corresponding to a given orientation angle are defined by the following relations:
\begin{equation}
\label{eq:lambda}
\begin{split}  
  \lambda & =\sqrt{L^2 \cos^2 \theta + D^2 \sin^2 \theta }, \\
  \delta & = L D / \lambda.
\end{split}
\end{equation}

These first of these expression is the standard result for the projection length of a rotated ellipse, the second ensures that the ellipse area is conserved under rotation. Since the focus of our analysis is on the impact of the geometry of many superimposed light-emitting structures on their collective emission pattern, we used a simplistic model for calculating the optical output of the individual structures. The structures were considered to be uniformly filled with a luminous plasma, with each pixel producing a unit dimensionless emission flux. Under this assumption, a LOS-integrated emission profile (the SPS ``pulse'') of a single elliptical structure such as the one illustrated in Figure \ref{fig_pulse}(a) is described by the function
\begin{equation}
\label{eq:pulse}
P_i(x) = \left\{
    \begin{array}{ll}
        \delta_i \, \sqrt{ 1 - \left[(x - x_{0i})(\lambda_i /2) \right]^2 },  &\, |x - x_{0i}| < \frac{\lambda_i}{2} \\
        0 & \mbox{otherwise,}
    \end{array}
\right.
\end{equation}
as shown in Figure \ref{fig_pulse}(b). It can be verified by combining Eqs. (\ref{eq:lambda}) and (\ref{eq:pulse}) that the total emission flux of the structure is equal to its cross-sectional area 
\begin{equation}
\label{eq:area}
s_i = \pi L_i D_i/4
\end{equation}
as expected, with the maximum emission level $\max\{P_i(x)\}=\delta_i$ reached at the projected center. 

A superposition of $N_p$ emission pulses, each described by its set of $L$, $D$, $\theta$, $x_0$ and $y_0$, defines the total LOS-integrated profile $I(x)$ of the emission model:
\begin{equation}
\label{eq:profile}
  I(x) = \sum_{i=1}^{N_p} P_i(x),
\end{equation}
where $i = 1,...,N_p$ is the index of the structure. The average filling factor of the model was calculated using
\begin{equation}
\label{eq:f}
  f = \frac{\sum_{i=1}^{N_p} s_i }{N_x \, N_y}.  
\end{equation}
To construct a model with a target $f$ value, randomly sampled structures were added to the simulation domain iteratively until the desired filling factor was reached. We found that this method is efficient for the filling factors of up to $\sim 0.5$, after which the computational cost of the non-overlapping constraint shows an explosive growth. 

Many previous observations suggest that the morphology of the thin solar corona involves light-emitting structures organized across wide ranges of spatial scales (see e.g. \citet{ gomez1993, uritsky2013, uritsky2014, aschwanden2016, mason2022, uritsky2023},  and refs therein). In closed-field coronal regions, this tendency manifests itself in the spatial clustering of small-scale loops and loop strands into multiscale loop bundles, possibly due to transient localized magnetic reconnection events resulting in clusters of hot flux tubes of different sizes \citep{knizhnik2018, knizhnik2020, klimchuk2023}. This clustering could also be consistent with the "coronal veil" hypothesis \citep{malanushenko2022} according to which an  apparent loop bundle could be produced by a single light-emitting structure with a complex geometry.  In many cases, the occurrence rate of multiscale coronal structures approximately follows  a power-law probability distribution. The range of scales demonstrating this behavior is typically limited from below by the size of the fundamental small-scale dissipative process heating the system, and from above by the characteristic length of the system as a whole.

To reflect this tendency on our model, we used a truncated power-law distribution
\begin{equation}
\label{eq:alpha}
p(L) = \left\{
    \begin{array}{ll}
        c\,L^{-\alpha},  &\, L \in [L_{\min}, L_{\max}] \\
        0 & \mbox{otherwise.}
    \end{array}
\right.    
\end{equation}
to sample the width of the SPS structures. Here, $\alpha \ne 1$ is the power-law distribution index, $L_{min}$ and $L_{max}$ are respectively the smallest and largest width of the superimposed structures, and $c$ is the normalization constant:
\begin{equation}
  c = \frac{1-\alpha}{L_{\max}^{1-\alpha} - L_{\min}^{1-\alpha}}.  
\end{equation}

The distribution index of luminous coronal structures has been evaluated in previous observational and numerical studies and it usually falls in the range $2 -3$. It should be noted that, in addition to generating scale-invariant sets of structures, the distribution model (Eq. \ref{eq:alpha}) can also be used to output structures with a fixed size by choosing $L_{min} \approx L_{max}$, or to sample the structures from the uniform distribution with the support $[L_{min},L_{max}]$ by setting $\alpha = 0$. 

The anisotropic shape of the structures was controlled by the additional power-law relation 
\begin{equation}
\label{eq:gamma}
    D(L) = \frac{D_{\min}}{L_{\min}^\gamma}\, L^\gamma
\end{equation}
in which $\gamma$ is the anisotropy index, $D_{min}$ is the smallest allowed thickness of the added structures, and the normalization ensures $D(L_{min})=D_{min}$. 
 
If $\gamma \ne 1$, Eq. (\ref{eq:gamma}) results in a scale-dependent behavior of the aspect ratio $L/D \propto L^{1-\gamma}$, which is consistent with previous theoretical investigations of multiscale current sheets in turbulent magnetohydrodynamic (MHD) plasmas \citep{uritsky2010}. The conditions $\gamma > 1$ ($\gamma < 1$) result in the anisotropy occurring predominately at small (large) spatial 
scales. \footnote{For $\gamma > 1$, the aspect ratio decreases monotonically with $L$ under the condition $L/D > 1$, which requires $L < L_{min}^{\gamma/(\gamma-1)} D_{min}^{1/(1-\gamma)}$; for smaller $L/D$ ratios, the width and the thickness of the structures switch roles leading to an ambiguity. The $\gamma > 1$ regime is not used in the present study.}
For $\gamma=1$, the aspect ratio becomes constant, and it takes the value 1 under the additional constraint $D_{min} = L_{min}^\gamma$ yielding perfectly circular structures with diameter $L=D$.

The described set of rules and equations makes it possible to generate a continuum of  mathematical models reproducing a variety of physical scenarios defining the cross-sectional morphology and the resulting emission profiles of the coronal loops. These models could be used to infer the fine geometry of the loop emission (which is usually inaccessible to the observer due LOS ambiguities in the optically thin corona) based on the properties of the directly observed LOS-integrated emission profiles, see Eq. (\ref{eq:profile}). In this paper, we study a small set of narrowly determined SPS models reproducing three archetypal cross-sectional geometries of a multistranded loop system, leaving a systematic investigation of a broader class of such models for future studies. 


\begin{deluxetable*}{ll|cccc}{h}
\label{table_models}
\tablecaption{Parameters of the three constructed SPS Models 1-3 and the MHD turbulence Model 4 (in all models, $N_x = N_y = 1024$).  \label{tab:models}}
\tablewidth{0pt}
\tablehead{
\colhead{Parameter} & \colhead{Interpretation} & \colhead{Model 1} & \colhead{Model 2} & \colhead{Model 3} & \colhead{Model 4} 
}
\decimalcolnumbers
\startdata
 $N_p$ & number of pulses & $4\times10^5$ & $2\times10^6$ & $6\times 10^6$ & $\sim  10^5$\\
 $N_s$ & number of slits & 1000 & 1000 & 1000 & 153 \\
 $f$ & filling factor & 0.30 & 0.30 & 0.30 & 0.02\\
 $L_{min}$ & minimum size & 20 & 3  & 3 & - \\
 $L_{max}$ & maximum size & 20 & 300  & 300 & -\\
 $D_{min}$ & minimum thickness & 20 & 3  & 2 & - \\
 $\alpha$ & distribution index & - & 2.5 & 2.5 & 2.2 \\
 $\gamma$ & anisotropy index & 1.0 & 1.0 &  0.4 & 0.2 \\ 
\enddata
\end{deluxetable*}


{\bf Model 1 (fixed-scale, isotropic)} mimics a simple scenario of a corona filled with circularly-shaped luminous structures of a constant diameter $L=20$ which are randomly spread across the emitting volume. Since the size of the structures is dimensionless, they could be associated with various physical morphologies such as e.g. single loop strands or thin multistranded loops embedded into a larger loop system.

{\bf Model 2 (multi-scale, isotropic)} represents a random superposition of circular emitting structures whose diameter is drawn from the power-law distribution (Eq. \ref{eq:alpha}) spanning over two orders of magnitude. The distribution index $\alpha$ is set to 2.5 to match the value of the average power-law exponent describing the linear size distribution of multiscale emitting structures in the extreme ultraviolet corona \citep{uritsky2013}. We note that in general, the value of $\alpha$ needed to simulate a specific loop system will be somewhat different from the one used in this SPS model, and it may vary in time depending on the local solar conditions.

{\bf Model 3 (multi-scale, anisotropic)} implements the most general rules of the three SPS models by allowing the structures to have random sizes and be also anisotropic and randomly oriented. The anisotropy index $\gamma = 0.4$ used in this model results in highly elongated loop cross-sections whose aspect ratio $L/D$ increases with size. The distribution index of this model is chosen to be 2.5 to simplify the comparison with Model 2.  

While the choice of the parameters used in the three SPS models described above is to some degree justified by solar observations, we would like to stress that fitting the measurements was not the goal of our runs. Rather, our main objective is to demonstrate a qualitative relationship between the geometry of the individual emitting structures and the resulting LOS-integrated emission signature, in the absence or presence of the multiscale statistical spread and the anisotropy of the structures. As shown below, these characteristics have a significant impact on the predicted emission profiles allowing one to narrow-down the scope of cross-sectional geometries consistent with the observed coronal loop systems.

\subsection{The MHD turbulence model}
\label{sec:mhd}

MHD turbulence, understood in a broad sense, is often invoked to explain complex plasma geometries appearing in the fluid range of scales, including highly fragmented current and vorticity sheets and/or filaments \citep{frisch1995}. MHD flows tend to generate strongly intermittent density structures. In the presence of strong axial magnetic field, these structures could naturally produce loop-like signatures such as the ones investigated by \citet{malanushenko2022}. Model 3 with $\gamma < 1$ can be considered as a toy model of this effect since the elongated structures oriented along the LOS are expected to result in strong and narrow emission spikes resembling optical signatures of LOS-aligned intermittent current sheets. To further test this scenario, we constructed an additional model, labeled below as Model 4, predicting optical signatures of an ensemble of highly intermittent structures produced by turbulence.

{\bf Model 4 (MHD turbulence)} is based on the data from an earlier numerical simulation of a decaying incompressible resistive three-dimensional MHD turbulence in the absence of an imposed uniform magnetic field \citep{mininni2006, uritsky2010}. The magnetic Prandtl number was taken equal to unity,  with the Taylor Reynolds number reaching 1100. The solution was obtained for a periodic $1536^3$ volume subject to the Arnol'd-Beltrami-Childress initial condition with a fully helical velocity and magnetic field. Due to the lack of an ambient magnetic field, the model is not well suited for studying the three dimensional configurations of coronal loops; however, its solutions could be relevant to analyzing the effects of turbulent intermittency on two-dimensional loop cross-sections.

For computing the LOS-integrated profiles of the MHD turbulence model, we used two-dimensional slices of the the square current density which was found to be a sensitive marker of spatial clustering in the model \citep{uritsky2010}. The slices were intended to statistically mimic cross sections of a complex loop system populated with multiscale strands and current sheets. As with the SPS models, the emission signal from the current density arrays was computed in a simplistic fashion, while placing the emphasis on the geometry of the emitting structures rather than the amount of the emission. The dimensionless emission output from a single pixel was taken equal to the local square current density, after which the LOS-integrated emission profile was constructed by summing up over local emission fluxes along one of the transverse directions. It should be noted that although the current density is a close proxy to the Ohmic heating in incompressible MHD, it does not account for several other first-principle energy conversion processes heating  the coronal plasma. In addition, \jak{the spatial distribution of coronal emissivity can be quite different from that of the heating rate for two reasons. First, coronal heating energy is spread rapidly along field lines. Second,} the emissivity is a complex function of local plasma conditions.  \vmu{In view of these limitations, we use Model 4 as a morphological model of possible projection effects caused by sheet-like luminous structures expected to be ubiquitous in a high-Reynolds number solar corona, leaving an analysis of more sophisticated physical models for future studies}. 

The three studied SPS models were defined on a rectangular grid with $N_x = N_y = 1024$, with the POS direction described by the $x$ coordinate. The runs were repeated $1000$ times to produce $N_s = 1000$ virtual slits used for performing statistical averaging described in the next section. The total number $N_p$ of superimposed pulses (Eq. \ref{eq:pulse}) included in each model was varied between $\sim 5 \times 10^5$ and $5 \times 10^6$ for obtaining comparable filling factors (Eq. \ref{eq:f}) which ranged between 0.23 and 0.30. The current density slices $j(x,y)$ were extracted from the three-dimensional $1536^3$ MHD model grid at a single time step representing a fully-developed state of the turbulent flow. The slices were separated by 10 grid nodes in the $z$ direction, and were rebinned to match the dimensions of the SPS models, resulting in a set of 153 virtual slits, each slit containing 1024 pixels. The LOS direction used to compute the optical emission was chosen to be parallel to the $x$ ($y$) axis for the odd (even) MHD slices to reduce statistical dependence between the emission profiles from the adjacent slices. 

The width $L$ of the cross section of the current density structures in the MHD model ranged between 5 and 200 pixel units of the rebinned grid; the crossover between the dissipative and inertial ranges was at $L \sim 30$ according to previous studies. The inertial - range $\alpha$ and $\gamma$ indices are respectively $2.2$ and $0.2$ \citep{uritsky2010}, and the average filling factor estimated using the structure detection method described in the above mentioned paper is about 0.02. 

The grid size of all numerical models is arbitrary and is not to be directly compared to the size of the observed coronal loop systems described later in the text. The pixel size is assumed to be larger than the width of the point spread function characterizing solar observations. See Table \ref{tab:models} for the summary of grid configurations and control parameters in each numerical model.

\subsection{Visual comparison}
\label{sec:demo}

\begin{figure*}
\begin{center}
\includegraphics[width=4.56 cm]{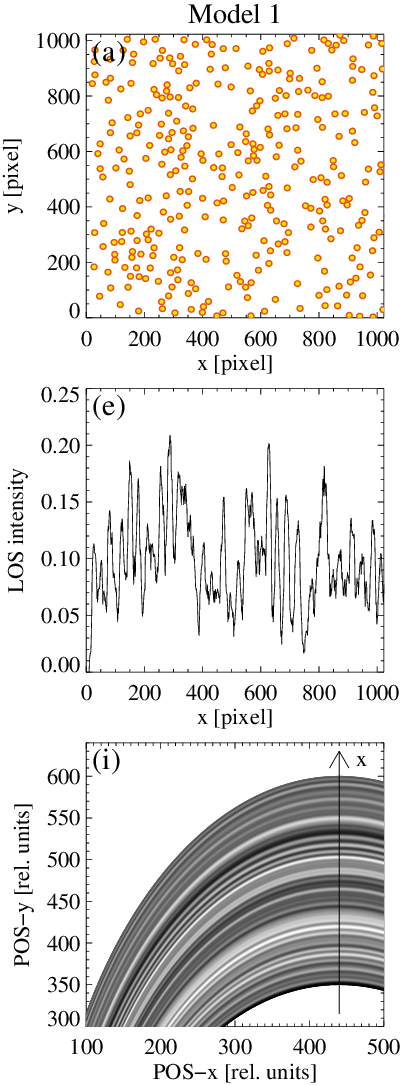}
\includegraphics[width=4.2 cm]{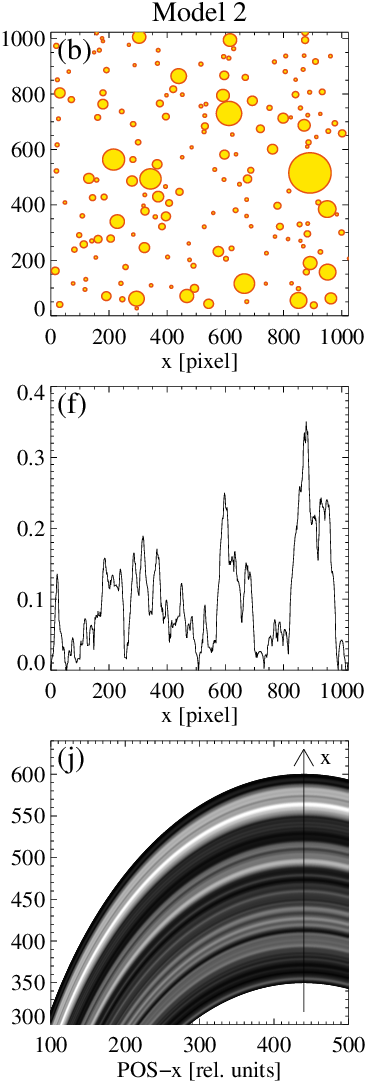}
\includegraphics[width=4.2 cm]{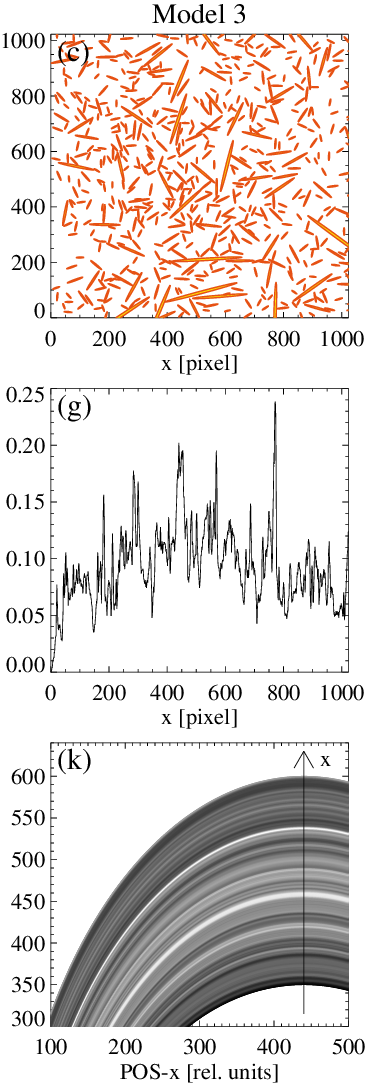}
\includegraphics[width=4.2 cm]{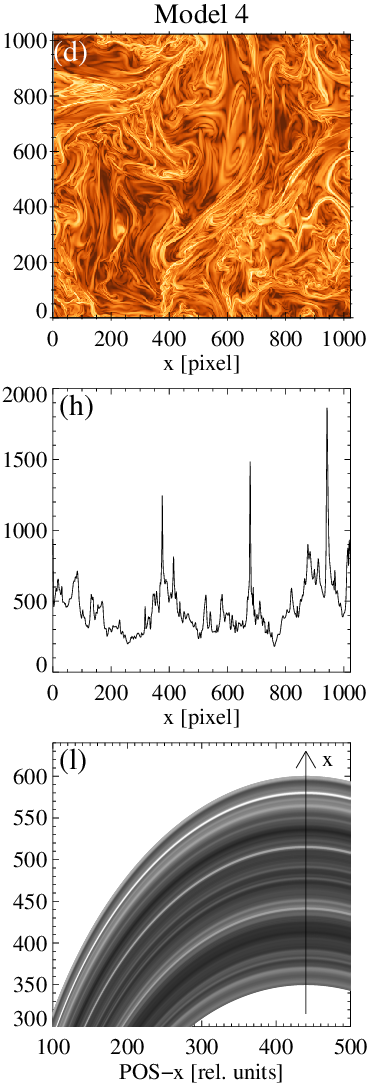}
\caption{\label{fig_model_demo} Visual comparison of typical cross-sectional geometries of the four  stochastic models of coronal loops studied in this paper: Model 1 composed of loop strands with the circular cross-section of a constant diameter ($L=D=20$), Model 2 containing power-law distributed circular strands ($\alpha = 2.5$, $\gamma = 1.0$), Model 3 with power-law distributed, randomly oriented strands with elliptical cross-sections ($\alpha = 2.5$, $\gamma = 0.4$), and Model 4 -- the MHD turbulence model based on a high-resolution decaying three-dimensional MHD turbulent simulation with the Arnol'd-Beltrami-Childress initial condition. Panels (a-d):  graphic representation of the cross-sectional pattern in each model. The shapes plotted on panels (a-c) are the synthetic loop strands in the three SPS models; the image on panel (d) is a 2D slice of the square current density in the MHD model, plotted on the logarithmic scale, with the white color corresponding to the most intense current. Panels (e-h): the LOS-integrated emission profiles obtained by integrating the simulated loop cross-section (a-d) along the $y$ axis. Panels (i-l): simulated plane-of-sky (POS) images of synthetic loop systems rendered using the obtained LOS emission profiles (e-h). The intensity scale of the synthetic loops is adjusted to the dynamic range of the corresponding profile signal, with the black (white) color representing the lowest (highest) LOS-integrated value. }
\end{center}
\end{figure*}

Figure \ref{fig_model_demo} demonstrates characteristic examples of simulated emission cross-sections (upper panels), the resulting LOS-integrated profiles (middle panels), and mathematically reconstructed side views of the coronal loop systems corresponding to each profile (bottom panels). The loop reconstruction was performed assuming that the shape of the cross-sectional emission profile is the same at all locations along the loop's artificially curved axis, and it is linearly rescaled to mimic the characteristic convergence of the magnetic field near the coronal base. To better illustrate the geometry of the SPS models, including the intensity profiles, their filling factors were reduced by a factor of 3 compared to the $f$-values (Table \ref{tab:models}) that were used for the quantitative analysis of the models.

LOS - integrated intensity profiles of Models 1 (fixed-size circular structures) and 2 (multiscale isotropic structures) plotted respectively on panels (e) and (f) of Figure \ref{fig_model_demo} represent samples of a colored stochastic process as discussed in the next section. The profile of Model 1 contains a well defined characteristic scale consistent with the diameter of a single structure, shows no significant trends at larger scales, and is smooth at the smaller spatial scales. Model 2 exhibits a more complex profile containing features on a variety of scale reflecting the multiscale nature of the underlying set of luminous structures. 

It is important to note that neither Model 1 nor Model 2 shows narrow isolated peaks characteristic of intermittent signals. These peaks are, however, clearly present in the LOS profiles of Model 3 (multiscale anisotropic structures, panel g), and are even more prominent in the MHD turbulence model (panel h). The intermittent peaks observed in these models are directly related with the anisotropy of the structures.  A close investigation reveals strong spatial correlation between the positions of the intense peaks on the LOS profiles of Model 3 (panel g) with narrow elongated structures oriented approximately in the LOS direction (panel c). The spikes in the MHD model's profile (panel h) seem to coincide either with the locations of compact thin current sheets possessing a relatively simple planar geometry and oriented along the LOS, or with the locations of sharp folds on larger current sheets which are also aligned with the integration direction, producing a strong and narrow emission signal. 

The difference in the shape of the LOS profiles of the the four models leads to a predictable difference in the visual appearance of the reconstructed loop images shown in the lower row of panels on Figure \ref{fig_model_demo}. The Model 1 loop (panel i) includes thin bright filaments of a similar width which are randomly spread over the loop system and are characterized by an approximately the same optical intensity. The loop produced by Model 2 exhibits a highly non-uniform transverse structure with no distinct characteristic spatial and intensity scales. The filaments that can be picked with unaided eye have drastically different a thickness, and are separated by wide dark regions with no significant filamentation. In contrast, Model 3 generates a multitude of narrow loop filaments with a well-defined apparent transverse size. This size is expected to be close to the minimum thickness $D_{min}$ of the underlying emitting structures included in the model. Unlike Model 1 whose loop filaments have similar sizes {\it and} optical intensities, Model 3 produces  detectable filaments characterized by a broad range of luminosities. This tendency is even stronger in the loop system synthesized based on the MHD turbulence profile (panel l). In that case, the reconstructed loop contains a small number of very thin, very intense insulated filaments with a varying brightness appearing against a dim  unstructured background. 

An experienced solar observer will likely quickly classify the loops generated by the fixed-size Model 1 and the MHD turbulence model as totally unrealistic, giving a chance to the other two models. Our quantitative analysis leads to the same general conclusion, ruling out the above-mentioned models and speaking in favor of a weakly anisotropic multiscale scenario represented by a statistical blend of Models 2 and 3.

\section{Data analysis methods and scaling relations}
\label{sec:methods}    

We employed two groups of statistical analysis methods characterizing different aspects of the irregular shape of the LOS-integrated emission profiles produced by the models and obtained from observations. Spectral analysis techniques were used to determine relative contribution of different spatial scales to the profile shape. Since spectral analysis alone is insufficient to characterize non-Gaussian fluctuations (see e.g. \citet{klimchuk2021}), we also employed higher-order statistical methods focusing on the investigation of profile gradients and spikes associated with intermittency. As shown below, a simultaneous analysis of spectral and intermittency measures enables an unambiguous classification of the studied signals. 

The ranges of spatial scales used for evaluating spectral and structure function exponents were chosen based on the following criteria. For the fixed-scale Model 1, the lowest analyzed scale was set at twice the diameter of the loop strands to focus the analysis on their mutual arrangement and not the shape of a single strand. The highest scale was chosen to be half the length of the LOS-integrated profile, which is the largest reliable scale. For the scale-invariant Models 2 and 3, these scales were set, respectively, at twice the $L$-size of the smallest strand ($2 \times L_{min} = 6$) and half the size of the largest strand ($L_{max} / 2 = 150$), in order to obtain a statistically representative picture of stochastic fluctuations formed by the multiscale pulses. Model 4 outputs were investigated across the interval of scales $[21, 210]$ \vmu{corresponding to current sheet structures generated by turbulent cascade  (see \citet{uritsky2010} for details). The shorter (dissipative) scales in real plasmas are controlled by ion-kinetic effects \citep{schekochihin2009, uritsky2014a} which are unresolved in solar observations and are therefore irrelevant to the purpose of this study.}

\subsection{Spectral analysis}
\label{sec:spectrum}

The slit-averaged Fourier power spectrum of a set of $N_s$ LOS-integrated emission profiles was defined as
\begin{equation}
\label{eq:spec}
  S(k) = \frac{1}{N_s}\sum_{j=1}^{N_s}{ S_j(k)}, 
\end{equation}
where $S_j(k)$ is the spectrum of the integrated profile, $j$ is the slit (profile) index, and $k$ is the transverse wave number in the direction across the loop system. As a simple approach for quantifying the hierarchy of spectral components over a range of $k$ covered by the signal, we used the power-law fit 
\begin{equation}
  S(k) \propto k^{-\beta}  
\end{equation}
in which the spectral index $\beta$ was evaluated over a range of spatial frequencies consistent with the size distribution of the underlying emitting structures, as discussed above.

\begin{figure*}
\begin{center}
\includegraphics[width=4.62 cm]{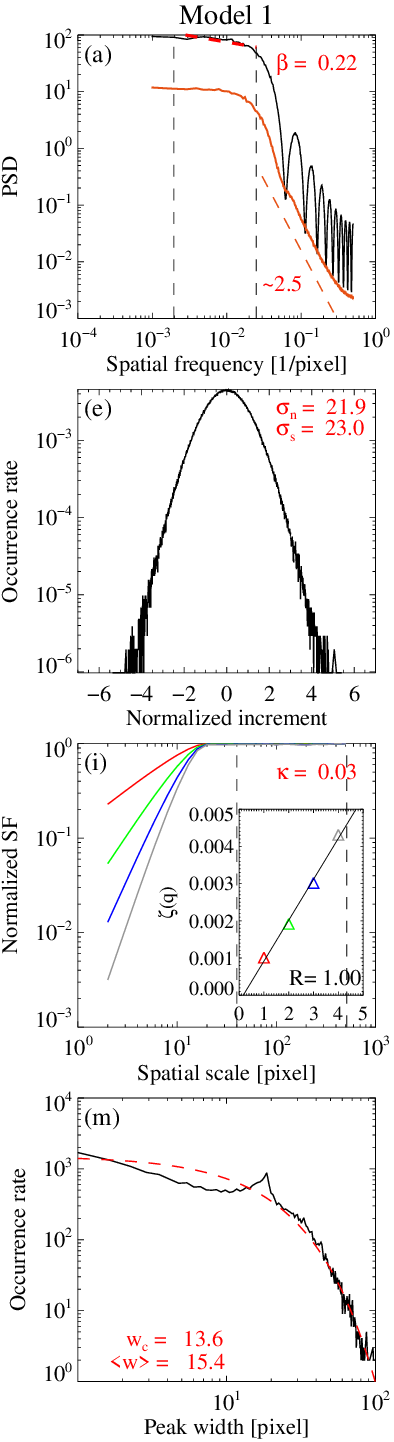}
\includegraphics[width=4.2 cm]{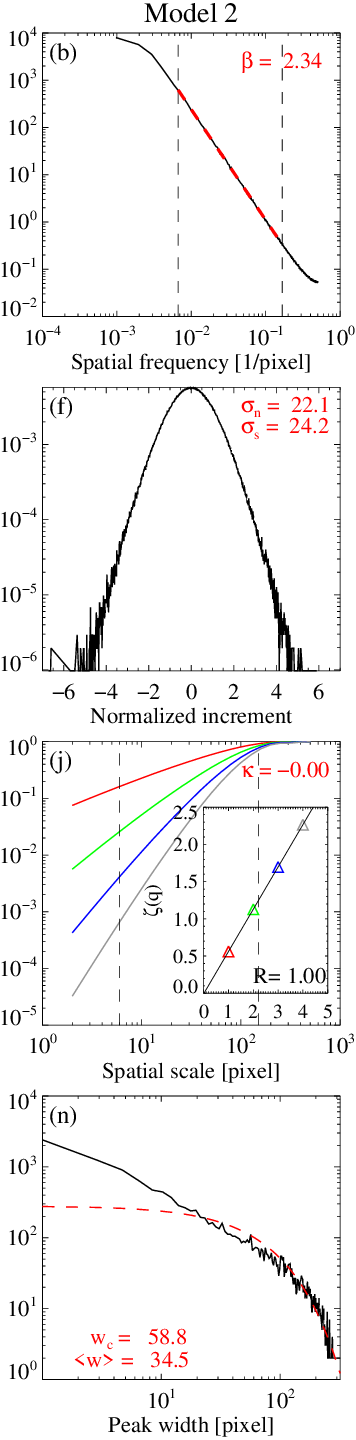}
\includegraphics[width=4.2 cm]{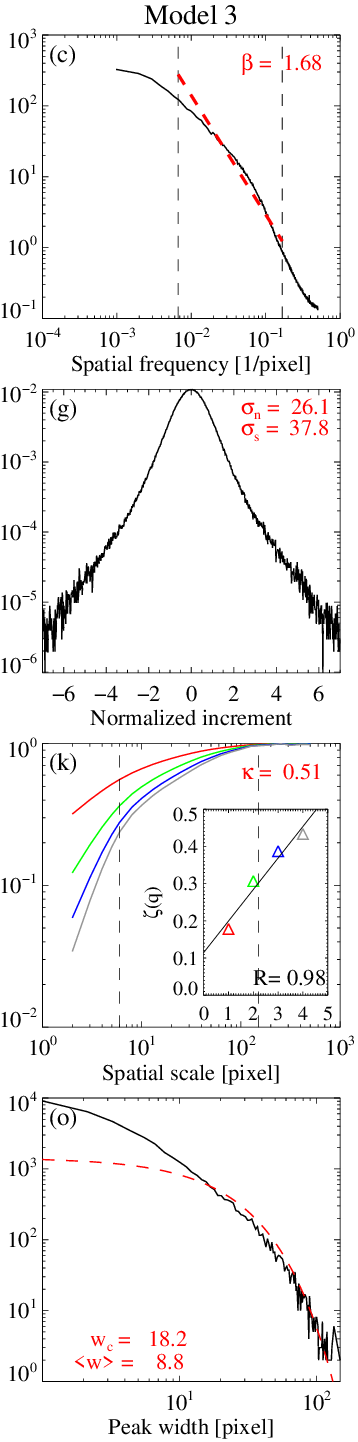}
\includegraphics[width=4.2 cm]{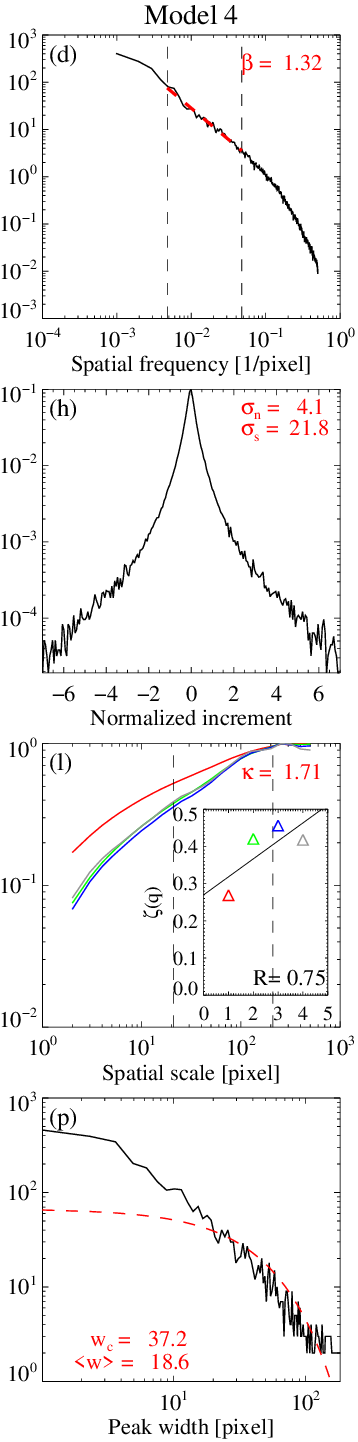}
\caption{\label{fig_model_stats} Statistical analysis of LOS-integrated profiles in SPS Models 1, 2 and 3 as well as Model 4 based on decaying MHD turbulence. Panels (a-d): energy Fourier spectra of the studied signals plotted as a function of the linear spatial frequency $k/2\pi$. Black dashed lines show the range of frequencies used to evaluate the power-law exponent $\beta$; red dashed lines represent the log-log slope given by the provided $\beta$ value. The orange curve added to panel (a) shows the spectrum of a randomized version of Model 1; the curve is shifted downward for easier comparison. Panels (e-h): probability histograms of the normalized 3-step increments of the LOS signals; $\sigma_n$ and $\sigma_s$ are respectively the normal and the sample-averaged standard deviations. Panels (i-l): Normalized structure functions of order $p=1$ (red), 2 (green), 3 (blue) and 4 (grey) versus the spatial scale $r$. The insets show the dependence of the SF exponent $\zeta$ computed over the range of scales shown by vertical dashed lines on the main plot on the SF order $p$, with $R$ being the Pearson correlation coefficient characterizing the linearity of the $\zeta(p)$ dependence. The provided values of the fourth-order Frisch intermittency index $\kappa$ evaluated using the extended self-similarity transformation as explained in the text. 
Panels (m-p): probability distributions of the peak width $w$. Solid black curve are the empirical histograms, dashed red lines are the exponential models described by the characteristic width $w_c$. Note the distinct peak at $w\approx 20$ on panel (m) matching the diameter $L$ of constant-size loop strands used in Model 1.}
\end{center}
\end{figure*}

The top row of panels in Figure \ref{fig_model_stats} shows the spectra computed for the studied stochastic loop models. The Model 1 spectrum shown with the black line on panel (a) is a combination of the oscillatory behavior
\begin{equation}
  E(k) \propto \text{sinc}^2(k \lambda_0 /2)
\end{equation}
owing to the discrete localization of the superimposed pulses of constant projected size $\lambda_0 = L_0 = 20$, and a power-law decay for $k > 2 \pi /\lambda_0$ reflecting the shape of the pulse. The oscillatory component becomes undetectable when the size of the structures is subject to a small statistical spread, as demonstrated by the second spectrum on the same panel (orange line) computed for a slightly modified version of Model 1, where  $L$ was drawn from a narrow normal distribution with a standard deviation equal to 5. The low-frequency part of the spectrum at $k < 2 \pi /\lambda_0$ is approximately constant for either version of the model, and is characterized by a near-zero average spectral index $\beta$. At larger wave numbers,  the spectral power decays approximately as $S(k) \propto k^{-2.5}$ as shown in the figure.

The spectra of the other three models shown in Figure \ref{fig_model_stats}(b-d) demonstrate a power-law decay with $\beta >0$ apparently resulting from the power-law distribution (Eq. \ref{eq:alpha}) of the structure sizes. To understand this connection, we extended the formalism developed earlier for a superposition of 1D exponential pulses \citep{schottky1926, jensen1989, kertesz1990, milotti2002} to describe 2D anisotropic structures with a POS footprint of an arbitrary form.

The spectrum of random pulses of characteristic size $\lambda$ representing POS projections of 2D structures of a cross-section area $s$ (see Eq. \ref{eq:area}) can be approximated by a generalized Lorentzian form 
\begin{equation}
\label{eq:lorentz}
E(k,\lambda) = \frac{s^2}{1 + (k \lambda)^b},
\end{equation}
combining a power-law decay $\sim k^{-b}$ at  $k> \lambda$ with a horizontal plateau at lower frequencies. This fit is consistent with the spectrum of the randomized version of Model 1, for which $\lambda = 20$ and $b\approx 2.5$, see Figure \ref{fig_model_stats}(a). The low-frequency plateau  represents the asymptotic solution for small $k$ under the normalization condition $E(k=0) = s^2$. The high-frequency power-law index $b \approx 2.5$ for circular structures as stated earlier (see Figure \ref{fig_model_stats}a), and it could take other values depending on the shape of the pulse after the LOS integration. The power spectral density of the emission profile (Eq. \ref{eq:profile}) is then given by the weighted sum of the individual contributions (Eq. \ref{eq:lorentz}):
\begin{equation}
\label{eq:E_int_1}
S(k) = \int_{s_{min}}^{s_{max}} p(s) \frac{s^2 }{1 + (k \lambda(s))^b} ds.
\end{equation}
Here, $s_{min}$ and $s_{max}$ are respectively the smallest and the largest emitting area in the superimposed set of structures, $p(s)$ is the density distribution of structures over area $s$, and $\lambda(s)$ is the characteristic pulse size corresponding to this area. Both functions need to be clarified before the integral given by Eq. (\ref{eq:E_int_1}) can be evaluated. 
Since  $s \sim LD$ and $D \sim L^\gamma$ (Eqs. \ref{eq:alpha} and \ref{eq:gamma}), the width of the structure should scale with its area as $L \sim s^\chi$, in which
\begin{equation}
\label{eq:chi}
\chi = \frac{1}{\gamma +1}.     
\end{equation}
Combined with Eq. \ref{eq:alpha}, this scaling requires that the area distribution $p(s)$ also  takes on a power law form: 
\begin{equation}
p(s) \sim s^{-\tau}.
\end{equation}
The power-law index $\tau$ of the area distribution can be calculated using the probability conservation condition $p(s)ds \sim p(L)dL$ by substituting $p(s) \sim s^{-\tau} \sim L^{-\tau/\chi}$, $ds \sim L^{(1-\chi)/\chi} dL$, and $p(L) \propto L^{-\alpha}$ (Eq. \ref{eq:alpha}). Solving the obtained scaling relation for $\tau$ and applying (Eq. \ref{eq:chi}), we find that
\begin{equation}
\label{eq:tau}
\tau = \chi (\gamma + \alpha) = \frac{\gamma+\alpha}{\gamma+1}.    
\end{equation} 

In its turn, the dependence $\lambda (s)$ takes different forms depending on whether the anisotropic structures have a constant or random orientation angle $\theta$. In the former case, $\lambda \sim s^{\chi}$ and the expression (\ref{eq:E_int_1}) becomes 
\begin{equation}
\label{eq:E_int_2}
\begin{split}
  S(k) & = \int_{s_{min}}^{s_{max}} \frac{s^{2-\tau}}{1+(k s^\chi)^b} ds \\ 
  & \propto  \frac{1}{k^{(3-\tau)/\chi}}\int_{k \lambda_{min}}^{k \lambda_{max}} \frac{u^{(3-\chi-\tau)/\chi}}{1+u^b} du,
\end{split}  
\end{equation}
where we used the substitution $s=(u/k)^{1/\chi}$. In the low-frequency limit, only the large-scale cut-off of the integral in Eq. (\ref{eq:E_int_2}) is important \citep{kertesz1990},  the integrated function reduces to  $\approx u^{ (3-\chi-\tau)/\chi - b}$, and the integral is convergent when the exponent is less than $-1$: 
\begin{equation}
  \frac{3-\chi-\tau}{\chi} - b < -1 
\end{equation}
 Taking into account Eqs. \ref{eq:chi} and \ref{eq:tau}, this convergence condition requires that $\alpha$ exceeds a critical threshold $\alpha_c$:
\begin{equation}
\label{eq:conv}
\alpha  > \alpha_c \, \equiv \, 2\gamma + 3 - b
\end{equation}

If the inequality (\ref{eq:conv}) is fulfilled, the scaling of the power spectrum is governed by the expression in front of the second integral in Eq. \ref{eq:E_int_2}: 
\begin{equation}
S(k) \propto  \frac{1}{k^{(3-\tau)/\chi}},
\end{equation}
and therefore the spectral power law index, after substituting expressions (\ref{eq:chi}) and (\ref{eq:tau}) one more time, can be expressed as
\begin{equation}
\label{eq:beta_theor}
\beta = \frac{3-\tau}{\chi} = 2 \gamma - \alpha + 3.
\end{equation}
This theoretical relation predicts that the spectral index should increase, yielding a steeper spectral decay, when the structures are more anisotropic (larger $\gamma$) and/or when their size distribution has a heavier tail (smaller $\alpha$). 

If the inequality (\ref{eq:conv}) is violated, the upper cut-off contributes to the dependence since the weight of the $\lambda \gg 1/k$ pulses becomes important: 
\begin{equation}
S(k) \propto  \frac{1}{k^{(3-\tau)/\chi}} k ^{(3 - \tau - \chi)/\chi - b +1 } = \frac{1}{k^{\,b}},
\end{equation}
and so the low-frequency behavior is always described by
\begin{equation}
\label{eq:beta_b}
\beta = b
\end{equation}
(with logarithmic correction for the special case $\alpha = \alpha_c$).  

The analytical results (Eqs. \ref{eq:beta_theor} and \ref{eq:beta_b}) are based on an asymptotic approximation which in practice requires that the size of the structures is distributed over many orders of magnitude. If the range of sizes is relatively narrow, and/or if the convergence condition is under question, Eq. \ref{eq:E_int_2} should be integrated numerically. 

\begin{figure}
\begin{center}
  \includegraphics[width=8.0 cm]{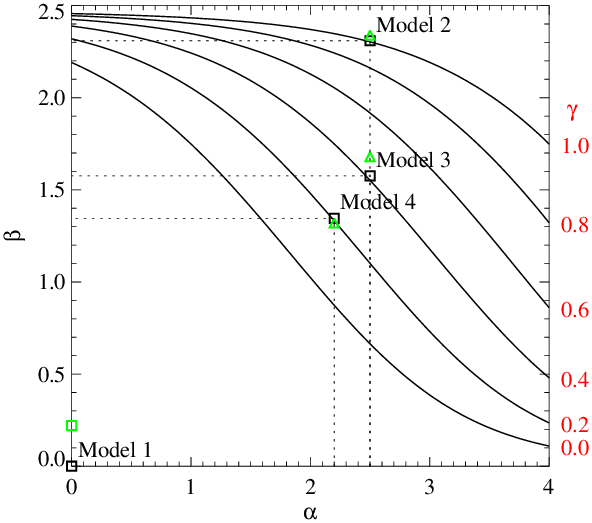}
  \caption{Theoretically predicted dependence of the spectral index $\beta$ of the emission profile on the distribution index $\alpha$ of the superimposed emitting structures, computed for a set of anisotropy indices $\gamma$. Expected and measured index values of the studied models are shown respectively with black squares and green triangles. }
  \label{fig_theor_beta}
\end{center}
\end{figure}

Figure \ref{fig_theor_beta} shows the predicted $\beta (\alpha, \gamma)$ dependence obtained using the numerical approach, with the integration limits representing the $L$-range used in the constructed models. The square symbols mark the spectral indices expected theoretically for each model; the green triangles show the measured values, according to Figure \ref{fig_model_stats}.

The power spectrum of Model 2 shown in Figure \ref{fig_model_stats}(b) exhibits a well-defined power-law form described by a single log-log slope $\beta \approx 2.3$, which matches the predicted value (Figure \ref{fig_theor_beta}) almost precisely. The spectrum of Model 3 has a more complex structure. The provided $\beta$ index is an average log-log slope obtained for the whole studied range of scales. However, strictly speaking, this spectrum cannot be fully described by a single $\beta$ value, as its local log-log slope varies between 1.0 to 2.3 depending on scale. This behavior results from the random rotation of the anisotropic structures constituting this model, and is not accounted for by the conducted theoretical analysis. Due to the rotations, the characteristic length scale (Eq. \ref{eq:lambda}) of the projected pulses becomes a function of two random variables, $L$ and $\theta$, and the maximum $\lambda$ consistent with the conducted scaling analysis becomes substantially smaller than $L_{max}$, which shifts the lower boundary of the wave numbers relevant to the theoretical results of this section toward higher values. The average index $\beta \approx 1.7$ of Model 3 is in a rough approximate agreement with the theoretical prediction ($\beta = 1.57$); the much smaller index observed at lower spatial frequencies is heavily affected by the rotational statistics which is beyond the scope of our current analysis. A more sophisticated treatment should include averaging over the two-dimensional probability density $p(s,\theta)$ when calculating the weighted sum (\ref{eq:E_int_1}); we leave this task for  future research.

The spectrum of the loop profiles produced by MHD turbulence (Model 4) can be reasonably approximated by a single power-law fit, with the spectral index $\beta \approx 1.3$ consistent with the prediction ($\beta = 1.35$) obtained for this model using previously published estimates of the inertial-range indices $\alpha$ and $\gamma$ \citep{uritsky2010}.  

\subsection{Intermittency analysis}
\label{sec:interm}

For a reliable classification of stochastic processes, spectral analysis should be complemented by additional methods addressing probabilistic aspects of the  process. The tendency of the rotated structures to produce sharp spikes in the LOS-integrated emission profile can be measured using data analysis methods focused on intermittency.

The simplest way to characterize the intermittency of the profile $I(x)$ is by studying the probability distribution of its two-point increments
\begin{equation}
\label{eq:incr}
  \Delta I(x, r) = I(x +r) - I(x),  
\end{equation}
where $r$ is the distance between the points. For non-intermittent Gaussian fluctuations, such as the Wiener stochastic process, fractional Brownian process, or fully developed homogeneous turbulence, the increments obey the normal distribution model for all values of $r$. A non-Gaussian probability density is typically associated with heavy tails of the  $\Delta I$ distribution (e.g. lognormal or power-law tails), which result in a non-vanishing occurrence rate of abrupt high-amplitude changes in $I(x)$ producing intermittent spikes. Departures from Gaussianity can be conveniently visualized by plotting the distribution of $\Delta I$ in semi-logarithmic coordinates transforming the normal distribution with zero mean and standard deviation $\sigma_n$ into an inverted parabola 
\begin{equation}
  \ln p(\Delta I) \propto  -\frac{1}{2 \sigma_n ^2} (\Delta I)^2,
\end{equation}
in which we made a simplifying assumption that the increments are centered around 0. The standard deviation was evaluated by measuring the width of the parabola \vmu{at the $p(0) / e $ level, where $e$ is the base of the natural logarithm.} It is expected to be lower than the sample-averaged standard deviation $\sigma_s$ in the presence of heavy non-Gaussian  tails. 

The second row of panels in Figure \ref{fig_model_stats} shows probability histograms of the three-point increments $\Delta I (x,3)$ in for the models. It is evident that the histograms of Models 1 and 2 (panels e and f, correspondingly) are very close to the Gaussian prediction, as confirmed by the parabolic shape of the curves and a close agreement between $\sigma_n$ and $\sigma_s$. The $\Delta I$ statistic of the isotropic multiscale Model 2 is  similar to that of the single-scale Model 1 despite the drastically different power spectra and cross-sectional geometries of the models. 

In contrast, the anisotropic Model 3 (panel g) demonstrates statistically significant departure from the normality expressed in heavy distribution tails and a larger discrepancy between the sample and normal standard deviation estimates compared to isotropic models. These signatures are consistent with the presence of intermittent spikes in the emission profiles produced by Model 3 (see Figure \ref{fig_model_demo}g). Model 4 based on MHD turbulence exhibits an even more substantial departure from normality, with  both the wings and the core of the histogram dominated by non-Gaussian effects, and the sample standard deviation $\sigma_s$ exceeding $\sigma_n$ by more than a factor of 5. The highly non-Gaussian statistics of increments in this model reflects its strongly intermittent behavior (Figure \ref{fig_model_demo}h) and is caused by thin current sheets self-consistently produced by  decaying nonlinear turbulence \citep{mininni2006, uritsky2010}. Since thin multiscale dissipative structures are inherent to any fully developed turbulent flow, one can expect the strongly non-Gaussian behavior of increments in Model 4 to be characteristic of turbulence in general.

Due to non-Gaussian effects, second-order statistics such as the Fourier power spectrum analysis are insufficient to quantify the intermittent signals. A more complete probabilistic testing is based on scale-dependent higher-order structure functions (SF) defined as the average increment (Eq. \ref{eq:incr}) raised to the power of $q$:
\begin{equation}
\label{eq:sf}
\begin{split}
  F_q(r) = & \left\langle | \Delta I(x, r) |^q \right\rangle_{x \in [0, N_x -r]} \\
         \equiv & \frac{1}{N_x-r} \sum_{x=0}^{N_x-r-1} | \Delta I(x, r) |^q. 
\end{split}
\end{equation}
Here, we use the unsigned definition of $F_q(r)$; the SF order $q \in \{1,2,3,4 \}$. For a multiscale intermittent process, the structure functions are approximated by a power law, with the power-law exponent $\zeta$ depending on the order $q$:
\begin{equation}
\label{eq:zeta}
  F_q(r) \propto r^{\,\zeta(q)}  
\end{equation}
To gain an intuitive understanding of this scaling ansatz, one can imagine a smooth analytical profile for which $\left\langle | \Delta I(x, r) |^q \right\rangle$ should be approximately proportional to $r^q$ for small enough $r$ at which the function can be linearized, and therefore $\zeta (q) \approx q$. A noisy signal with uncorrelated increments should produce the same $F_q$ on all scales independently on $q$, so $\zeta \equiv 0$. Other stochastic signals lie between these two extremes and are described by a set of $\zeta$ exponents that would generally increase with $q$ in either a linear or a nonlinear fashion (see e.g. \citet{abramenko2008, uritsky2017} and refs therein). 

The nonlinearity of the $\zeta(q)$ is considered to be a sensitive marker of intermittency. In many applications, it can be adequately described by a single parameter characterizing the $r$-dependence of the ratio of the structure function of the highest studied order $q_{max}$ to the lowest-order ($q_{min}$) function raised to $q_{max}/q_{min}$ \citep{frisch1995}. In our analysis, $q_{min} = 1$, $q_{max} = 4$, and the described ratio reads
\begin{equation}
\label{eq:kappa}
\frac{F_4(r)}{[F_1(r)]^4} \propto r^{-\kappa},
\end{equation}
where $\kappa$ is the Frisch intermittency index. Using Eq. \ref{eq:zeta}, it is easy to see that $\kappa = \zeta(4) - 4\,\zeta(1)$. For nonintermittent stochastic signals $\zeta \propto q$ and so $\kappa = 0$. For intermittent signals, the relationship between $\zeta$ and $q$ is nonlinear resulting in $\kappa > 0$, with the value of the index increasing as the intermittent spikes become higher and steeper.

The third row of panels in Figure \ref{fig_model_stats} shows the results of the structure function analysis of the four stochastic models. The structure functions of different orders are normalized to the maximum values and are color-coded for easier comparison. The insets show the estimated dependence $\zeta(q)$ and the results of its linear fitting characterized by the Pearson coefficient $R$, with the intermittency index $\kappa$ provided in each case.

The structure functions of Model 1 (panel i) show a power-law dependence on $r$ for the short scales below the deterministic structure size $L_0=20$, and reach a plateau at larger scales since the profiles contain no large-scale trends. The linear hierarchy of $\zeta$ exponents in the small-scale regime reveals a clear-cut non-intermittent behavior described by $R=1$ and a near-zero intermittency index $\kappa$. Model 2 (panel j) exhibits a substantially broader range of the $F_q(r)$ scaling reflecting the multiscale nature of the model. Nevertheless, its intermittency index and the Pearson coefficient indicate a complete lack of intermittent spikes in the LOS-integrated profiles. 

Structure functions of Model 3 (Figure \ref{fig_model_stats}k), as expected from its spiky LOS profiles and the non-Gaussian $p(\Delta I)$ distribution, exhibits unambiguous signatures of intermittency which include a statistically significant nonlinearity of the $\zeta(q)$ dependence and an elevated intermittency index $\kappa \approx 0.5$. A faster than a power law decline of the structure functions of Model 3 at the smallest scales is presumably caused by the scale-dependent aspect ratio $L/D \sim L^{(1-\gamma)}$ of the superimposed structures. As in the case with the increment analysis, Model 4 demonstrates the most pronounced signs of intermittency characterized by a high $\kappa \approx 1.7$ and a significant departure of $\zeta(q)$ from the linear prediction. 

\begin{figure}
\begin{center}
\includegraphics[width=8.5 cm]{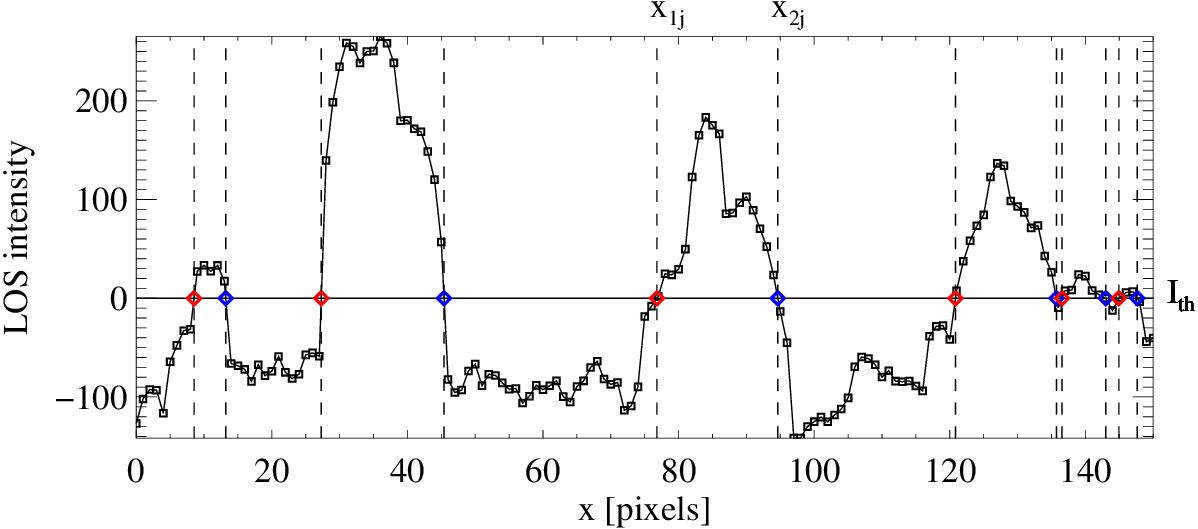}
\caption{\label{fig_threshold} Example of the threshold-based identification of local emission peaks in a detrended LOS emission profile obtained from a loop model or from a virtual solar image slit.  $I_{th}$ is the detection threshold, red and blue dots mark respectively the starting ($x_{1j}$) and ending ($x_{2j}$) positions of each detected peak, and the peak width $w_j = x_{2j} - x_{1j}$.} 
\end{center}
\end{figure}

The last method included in our intermittent analysis toolkit is based on the width statistics of the intermittent spikes. The width $w_j$ of the local emission  peak labeled with index $j = 1,...,M$, where $M$ is the number of the detected peaks, was estimated by applying a constant detection threshold set to the mean value of the signal, and computing the distances $x_{2j} - x_{1j} = w_j$ between the footpoints of each peak at the threshold level, as illustrated by Figure \ref{fig_threshold}. To reduce the contamination of the peak statistics by long-range correlations characterizing system-level geometry of the loop system, the profiles were detrended by subtracting second-order polynomial trend before identifying the peaks. Similar methods, typically employing more sophisticated edge detection algorithms, are used for measuring the width of the loop filaments and strands in the observed  coronal loop systems (see e.g. \citet{klimchuk2015, klimchuk2020}. We invoke this approach to quantify the multiscale statistics of the peaks. If the superposition is dominated by spikes of a predefined size (which is to be expected if the underlying structures have a narrowly-distributed projected $\lambda$), the probability distribution of spike width should exhibit an exponential decay for $w > w_c$:
\begin{equation}
\label{eq:peaks}
  p(w) \propto \exp{(-w/w_c)},  
\end{equation}
in which $w_c$ is the characteristic width of the spikes. Eq. \ref{eq:peaks} is a classical result for the asymptotic statistics of level-crossing intervals of short-range correlated Gaussian processes including those described by Lorentzian spectra (Eq. \ref{eq:lorentz}), see e.g. \citet{horsthemke2006} and refs. therein. The $w_c$ value obtained from the exponential model should be close to the sample-averaged width $\left \langle w \right \rangle$. For a broadband $p(\lambda)$ distribution, the exponential fit will be only approximately correct at the largest scales revealing the presence of the large-scale cutoff $L_{max}$, and the sample mean width should be substantially larger than $w_c$.

The histogram for the $p(w)$ distributions of the four stochastic models are presented in the bottom row of panels in Figure \ref{fig_model_stats}. It can bee seen that spike width distribution of Model 1 is reasonably close to the exponential fit, resulting in $w_c \approx \left \langle w \right \rangle$. It is also easy to notice that the characteristic size of the structures constituting the model has penetrated into the peak width statistics, producing a sharp local maximum at $w = L_0$. This suggests that a real-life loop system mimicked by this mono-scale model would also exhibit a distribution maximum at the spatial scale corresponding to the characteristic width of the superimposed loop filaments, enabling an evaluation of this important parameter based on the analysis of LOS-integrated profiles. This possibility is clearly off the table for the multiscale emission patterns represented by Models 2-4 whose $p(w)$ histogram, predictably, contains no statistically significant humps or peaks and shows drastic departure from the exponential fit, with the $\left \langle w \right \rangle / w_c$ ratio being substantially greater than 1 and roughly the same for the three multiscale models.

\section{Solar data analysis}
\label{sec:Hi-C}

\begin{figure*}
\begin{center}
\includegraphics[width=17.5 cm]{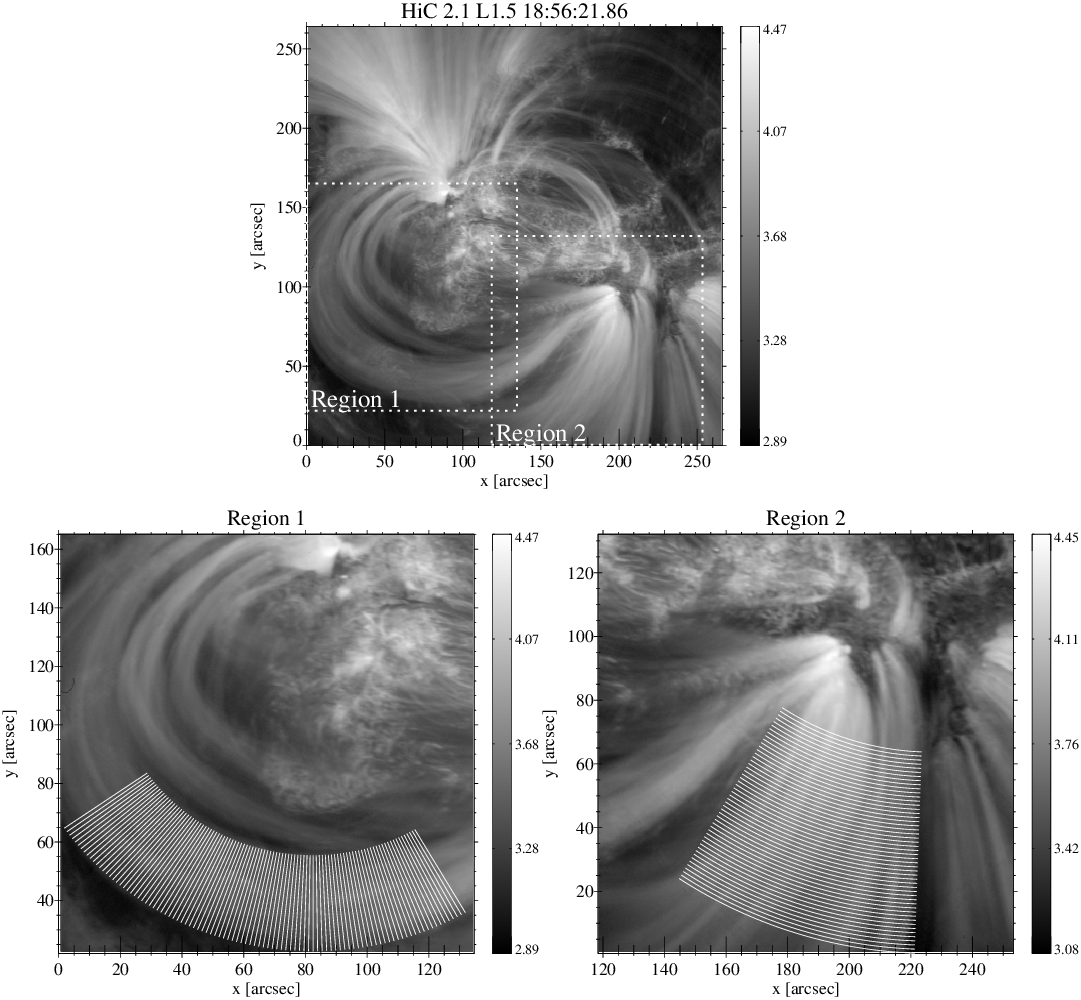}
\caption{\label{fig_hic_large} Two studied loop systems observed by Hi-C in the active region AR 12712. Top panel: the entire Hi-C field of view showing the loop systems, bounded by the dotted lines. Bottom left: the loop system of Region 1 with approximately parallel strands. Bottom right: the loop system of Region 2 exhibiting a significant expansion with height. White solid lines show the positions of the studied transverse slits. Radial and circular slit geometries were used to match the large-scale shape of the Region 1 and Region 2 loop systems, correspondingly. The slit positions displayed here were used for automatically processing all available stable images with relatively small motion blur obtained during the Hi-C 2.1 flight \citep{rachmeler2019}, resulting in 3600 slits for the Region 1 and 1800 slits for the Region 2 loops. Logarithmic intensities are plotted on all three panels to emphasize low-intensity features.}
\end{center}
\end{figure*}

\begin{figure}
\begin{center}
\includegraphics[width=8.0 cm]{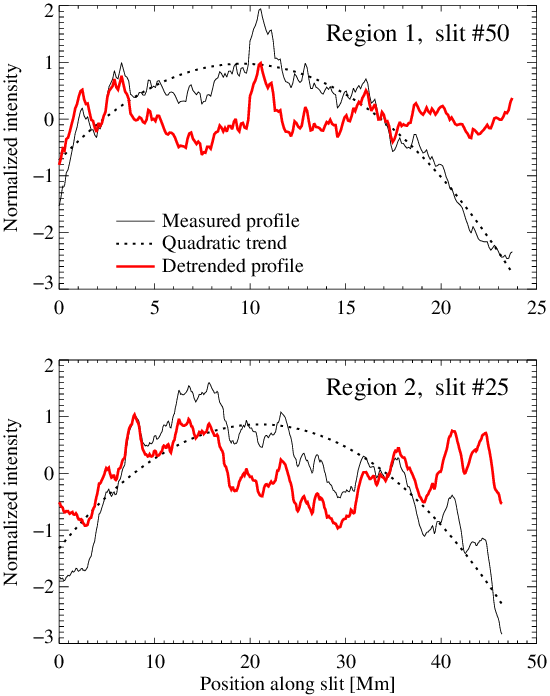} 
\caption{\label{fig_HiC_profiles} Examples of polynomial detrending of the Hi-C emission profiles intended to reduce the large-scale distortions caused by  limited cross-sectional sizes of the studied loop systems. The peak width statistics reported in Figure \ref{fig_HiC_stats}(g, h) were obtained from the detrended emission profiles (shown with red lines in the above examples).}
\end{center}
\end{figure}

\begin{figure*}
\begin{center}
\includegraphics[width=4.62 cm]{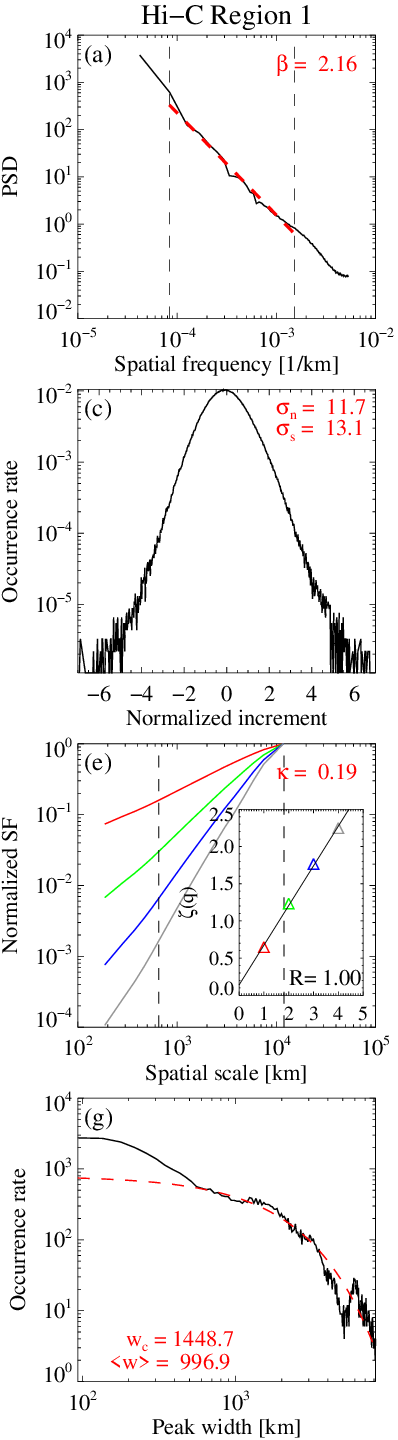}  
\hspace{0.5 cm}
\includegraphics[width=4.2 cm]{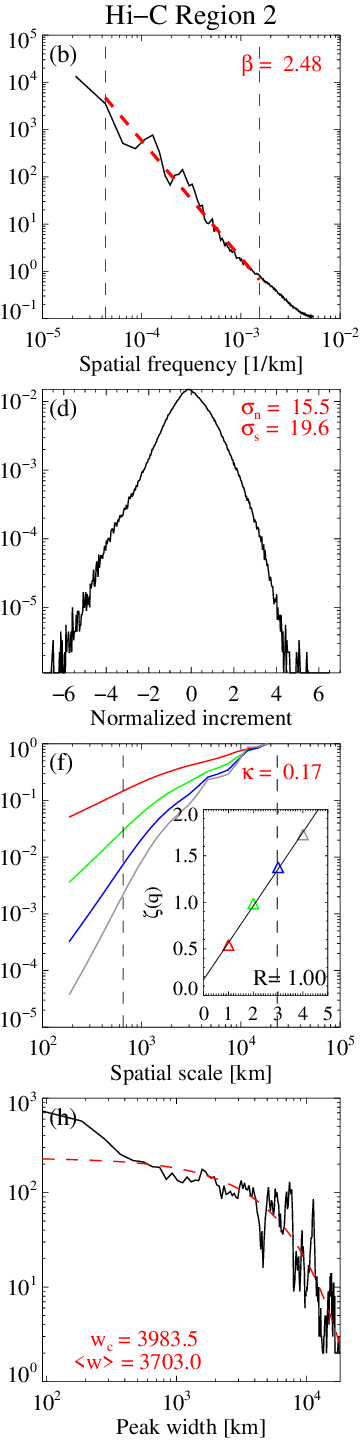}  
\caption{\label{fig_HiC_stats} Statistical analysis of LOS-integrated profiles extracted from the Hi-C images using the two systems of virtual slits shown in Figure \ref{fig_hic_large}. The statistical methods and notations are the same as in Figure \ref{fig_model_stats}, and are described in Section \ref{sec:methods}.}
\end{center}
\end{figure*}

We studied the images obtained by EUV telescope onboard the High-Resolution Coronal Imager (Hi-C) sounding rocket mission during its second successful flight which was performed on May 29, 2018 (flight Hi-C 2.1 \citep{rachmeler2019}. The target of the observation was NOAA Active Region (AR) 12712. About 80 images with the time cadence of 4.4 s were recorded during the 5 and a half minute observing time. The EUV telescope has an unprecedented plate scale of 0.129 arcsec/pixel and the time cadence of 4.4 s for targeting small-scale transient coronal features. The quality of the final image product was affected by a motion blur from rocket jitter reducing the effective spatial resolution. The low-blur images contain resolved coronal features of a linear size of at least 0.47 arcsec, and provide rich information about the cross-sectional shape of the observed  coronal loops \citep{williams2020a, williams2020b,  williams2021}.

For characterizing  the spectral and intermittency statistics of the loop cross-sections, we selected 36 high-quality Level 1.5 Hi-C 2.1 images labeled as ``low jitter'' frames in the Hi-C database \footnote{\vmu{The Hi-C 2.1 images and the low-jitter image flags were downloaded from \url{https://msfc.virtualsolar.org/Hi-C2.1}}}. The top panel of Figure \ref{fig_hic_large} provides an example of such an image. To compare cross-sectional structures in the absence and presence of  significant geometric expansion, two separate loop systems contained in the regions of interest shown in the Figure were investigated. The loop system of Region 1 is composed of closed strands with approximately the same radius of curvature and height. The  second loop system featured in Region 2 quickly expands with distance from the footpoints. These fanning loops also belong to a closed flux system as evidenced by the contextual whole-disk images \citep{rachmeler2019}. 

To collect sufficient statistics comparable with that obtained from the stochastic model, extended sets of virtual slits were constructed for each loop system as shown in the bottom panels of Figure \ref{fig_hic_large}, with the geometry of the spatial slit domains approximating the large-scale geometry of each loop. The Region 1 loop system was sampled using 100 radially diverging straight-line slits arranged into a toroidal configuration, each slit containing 255 pixels. The Region 2 loops were sampled using 50 arc-shaped, 500 pixel long slits creating an expanding wedge near the loop footpoints. In either region, slits were approximately perpendicular to the respective loop system. To reduce the influence of the large-scale geometry of the loop system on the statistics of spikes (Eq. \ref{eq:peaks}) representing fine loop filaments, the image profiles extracted using each virtual slit were detrended by subtracting a second-order polynomial fit from the originally measured profile (Figure \ref{fig_HiC_profiles}). As mentioned earlier, the same detrending procedure was applied for identifying spikes in the model data. The described procedure was repeated for all 36 image frames, resulting in $N_s=3600$ and $N_s=1800$ detrended cross-sectional profiles for  Region 1 and Region 2 loops, correspondingly. The average distance between the pixels along the slits was about 93 km. The spectral and  structure function analyses of the Hi-C profiles were performed over the entire range of spatial scales covered by the loop profiles, excluding the smallest scales contaminated by pixel noise and rocket jitter, and the largest scales not properly represented in the statistics.

Figure \ref{fig_HiC_stats} shows the results of the analysis of the two Hi-C regions defined above. It can be seen that the Fourier energy spectrum of both loop systems follows  an approximate power-law over a wide range of spatial frequencies. In addition to this stochastic background, the Region 2 spectrum (panel (b)) contains a set of harmonic components which could reflect the quasi-periodically arranged loop bundles seen in that region (Figure \ref{fig_hic_large}). The spectral indices describing the two regions are substantially different, with the higher $\beta$ of Region 2 representing the more structured large-scale loop morphology near the loop footpoints. For either region, $\beta \ge 2$.

The distribution of signal increments in Region 1 plotted on Figure \ref{fig_HiC_stats}(c) is close to an inverse parabolic shape expected from a Gaussian distribution in semi-logarithmic coordinates. The Region 2 distribution (panel {d}) is somewhat asymmetric, possibly due to the clustering effects inside the loop system in that region which were mentioned above, and on average is also not far from the normal distribution as evidenced by the sample standard deviation which is reasonably close to the normal prediction. 

The scaling of the higher-order structure functions, presented in panels (e,f) of Figure \ref{fig_HiC_stats}, is roughly consistent with the power-law model (Eq. \ref{eq:sf}), with the structure function exponents growing linearly with order $q$ ($R=1.0$) as predicted for non-intermittent stochastic models. The Frisch index, $\kappa$, is statistically significantly greater than zero but is low compared to Models 3 and 4, leaving a possibility of a weak intermittency. This interpretation is also confirmed by a small departure of the peak width distributions from the exponential predictions at the small spatial scales, contributing to a small but measurable discrepancy between the observed and predicted characteristic width (Figure \ref{fig_HiC_stats} (g, h)).

\section{Conclusions}
\label{sec:conclusions}

\begin{deluxetable*}{ll|ccccc|cc}
\tablecaption{Summary of studied scaling ranges and measured statistical parameters of the simulated and observed coronal loops \label{tab:stats}}
\tablewidth{0pt}
\tablehead{
\colhead{Parameter} & \colhead{Interpretation} & \colhead{Model 1} & \colhead{Model 2} & \colhead{Model 2a} & \colhead{Model 3} & \colhead{Model 4} & \colhead{Hi-C, Region 1} & \colhead{Hi-C, Region 2} 
}
\decimalcolnumbers
\startdata
   $\beta$ & spectral index & 0.22 & 2.34 & 2.60 & 1.68$^*$ & 1.32 & 2.16 & 2.48 \\
   $\kappa$ & intermittency & 0.03 & 0.00 & 0.10 & 0.51 & 1.71 & 0.19 & 0.17 \\
   $1 - R$ & nonlinearity & 0.00 & 0.00 & 0.00 & 0.02 & 0.25  & 0.00 & 0.00 \\
   $\sigma_s/\sigma_n - 1$ & non-Gaussianity & 0.05 & 0.10 & 0.37 & 0.45 & 4.32 & 0.12 & 0.26 \\
   $w_c /\left\langle w \right\rangle -1$ & multiscaling & -0.12 & 0.70 & 0.18 & 1.07 & 1.00  & 0.45 & 0.08 \\
\enddata

\vmu{$^*${\footnotesize The local spectral index of Model 3 varies between 1.0 to 2.3 depending on scale, see section \ref{sec:spectrum} for details.}}

$^{\,}$ 

\end{deluxetable*}

We have conducted an in-depth numerical investigation of three-dimensional projection effects which could influence the observed loop-like structures in an optically thin solar corona. Several archetypal emitting geometries have been tested, including collections of luminous  structures with circular cross-sections of fixed and random size, light-emitting structures with highly anisotropic cross-sections, and stochastic current sheets generated by MHD turbulence. A comprehensive set of statistical signatures has been employed to quantitatively compare the LOS-integrated emission signals predicted by these numerical models with the transverse loop profiles observed by the Hi-C instrument.

The two overarching questions of this investigation are (a) what kind of three-dimensional emitting geometry is most likely to result in the observed coronal loop profiles, and (b) whether or not the observed one-dimensional loops can in fact be projections of higher-dimensional structures such as current sheets and/or current sheet folds oriented along the LOS \citep{malanushenko2022}. The results obtained enable us to address both questions.

As shown in Section \ref{sec:Hi-C}, the Hi-C loop profiles exhibit a pattern of parameters suggesting an underlying emission geometry characterized by power-law scaling with a fairly small amount of intermittency. To place this conclusion in the context of the constructed loop models (Section \ref{sec:methods}), Table \ref{tab:stats} summarizes the predicted and observed values of the $\beta$ and $\kappa$ indices as well as several derived indices ($1-R$,  $\sigma_s/\sigma_n - 1$ and $w_c /\left\langle w \right\rangle -1$) providing quantitative measures of intermittency and non-Gaussianity. The cross-comparison between the six sets of values shown in the table indicates that the SPS Model 1 is drastically inconsistent with the Hi-C profile shapes since it predicts a near-zero spectral index, while the real loops are characterized by $\beta >2$. 

The spectral indices of Model 3 and the MHD turbulence Model 4 are greater than zero but are still significantly lower that the $\beta$ values of the observed loops across most of the studied scales. Even more importantly, these highly anisotropic models demonstrate unrealistically high levels of intermittency and non-Gaussianity compared to the corresponding levels observed for the Region 1 and 2 loop systems. Based on the values listed in the Table, Model 2 is the closest of the four studied models to the coronal loops imaged by Hi-C. Its most important difference from the real-life loops is captured by the intermittency index $\kappa$ with is, by definition, close to zero in the model and is distinct from zero in the Hi-C data.

Provided that the studied coronal loops are statistically representative, the results of our analysis strongly suggest that fine transverse structure of the corona cannot result from a LOS overlap of fully isotropic, fixed-size emitting substructures, since they predict a near-zero spectral exponent contradicting the observations. Neither can it result from a superposition of multiscale, strongly anisotropic structures such as turbulence-generated thin current sheets, since such structures would cause an unrealistically strong intermittency in the LOS emission profiles unsupported by the observations. We note that the first of these conclusions, when extrapolated to sub-resolution scales, rules out a possibility that the loop can be composed of fine strands of approximately the same width. The second conclusion questions the scenario in which the observed loops are typically caused by kinks and folds in a thin corrugated quasi-two dimensional emitting structure (the coronal ``veil'', see \citet{malanushenko2022}) since such folds would, on average, produce a much higher level of intermittency in the cross-sectional loop profiles than the one seen in the data. 

\begin{figure*}
\begin{center}
\includegraphics[width=6.8 cm]{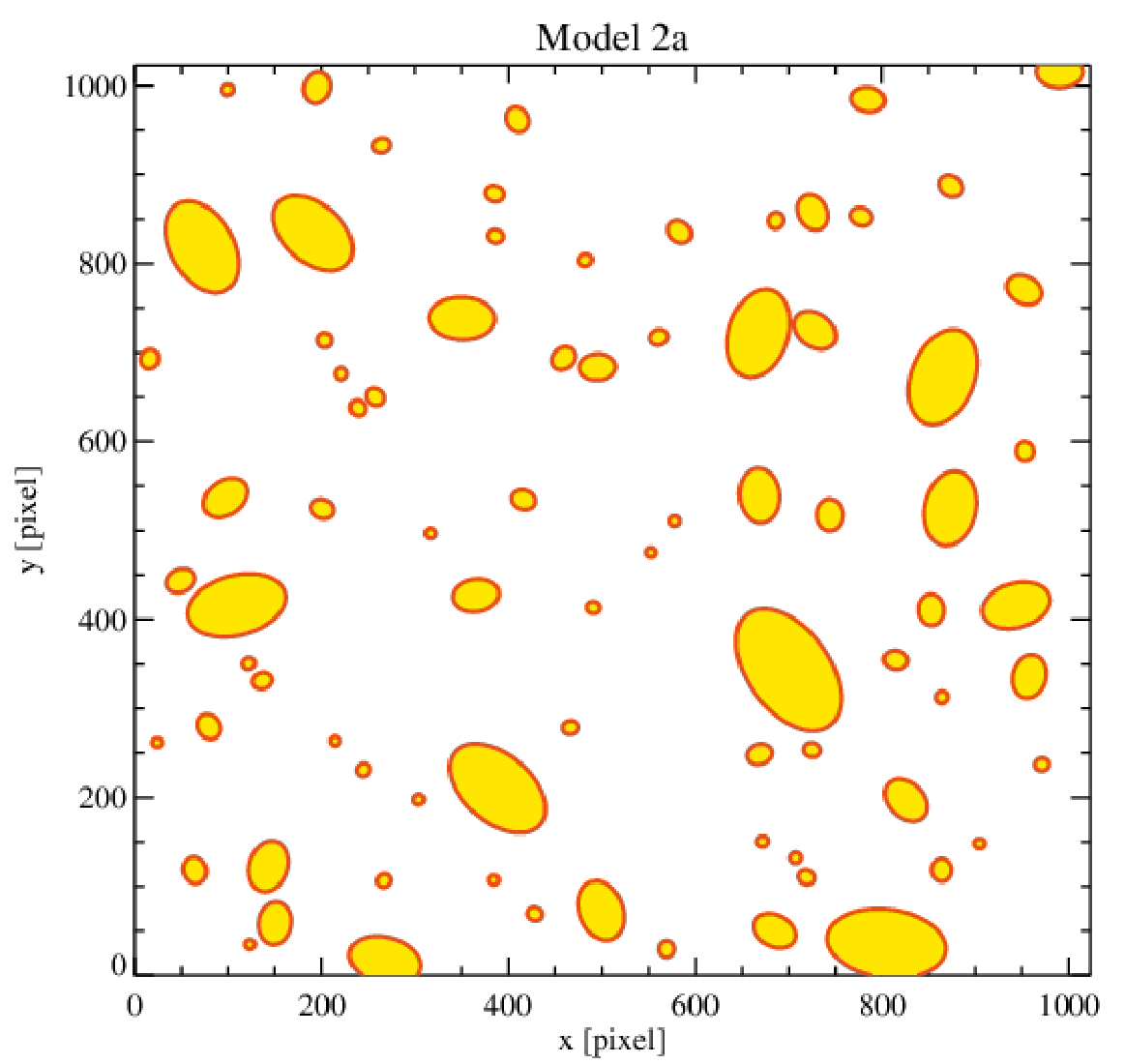} \hspace{1cm}
\includegraphics[width=8.0 cm]{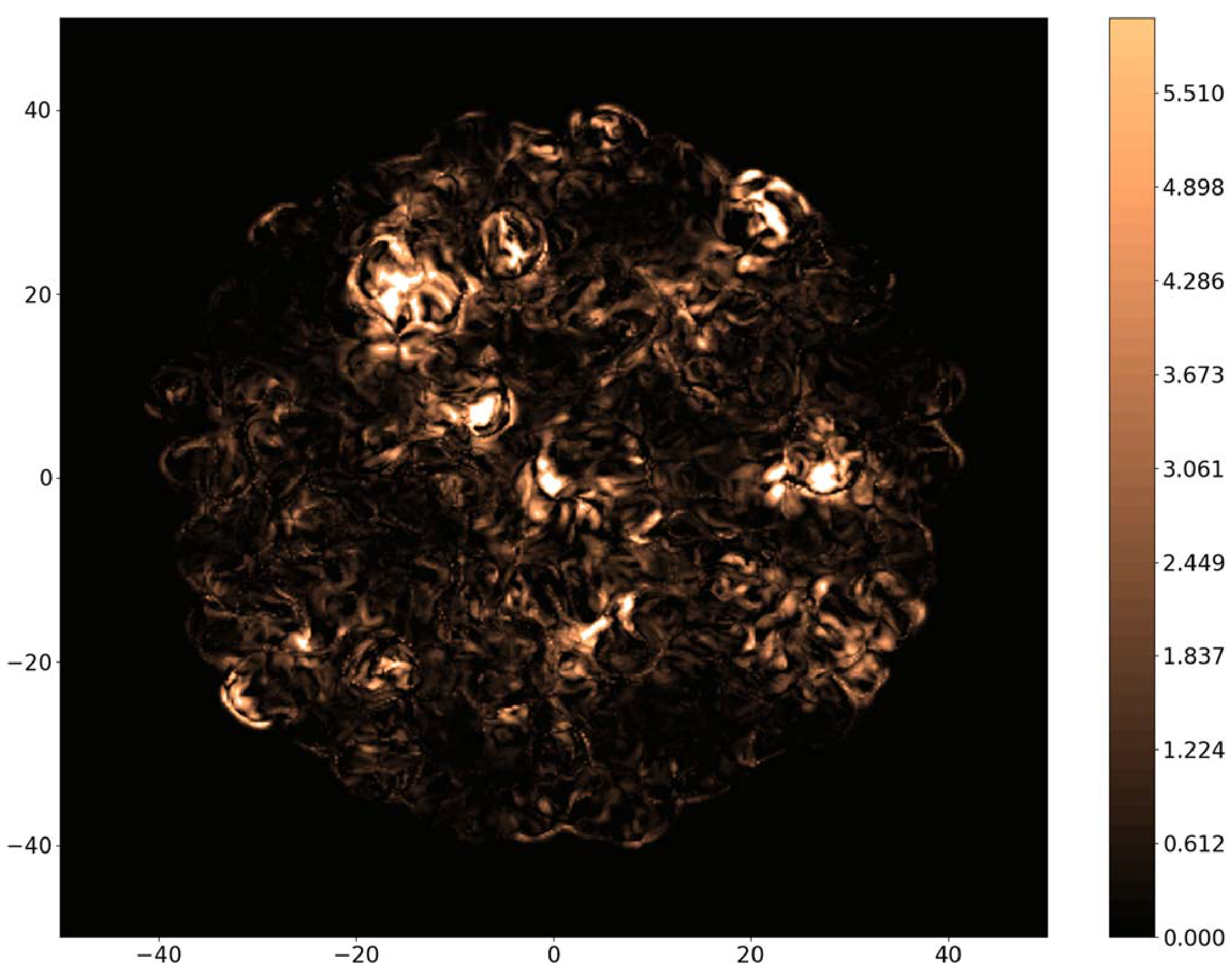}
\caption{\label{fig_nanoflares} 
Qualitative comparison of the cross-sectional loop geometry of the tuned-up SPS Model 2a (left) and  a 193{\AA} synthetic emissivity map  (right) produced by the MHD nanoflare model \citep{knizhnik2018,klimchuk2023}.
The parameters of the SPS model ($\alpha=1.7$, $\gamma=0.8$) produce LOS profile statistics that are roughly consistent with the Hi-C 2.1 observations.}
\end{center}
\end{figure*}

The best candidate to fitting the data is a scale-invariant SPS model, such as Model 2, with a small amount of added anisotropy. One possible physical scenario consistent with this mathematical model is a driven multi-stranded corona heated by clusters of nanoflare events of various sizes, as demonstrated in recent MHD simulations \citep{knizhnik2018, knizhnik2020, klimchuk2023}. The aspect ratio of these clusters could be different from 1 because of anisotropic driving conditions imposed by the photospheric flow. Alternatively, if the nanoflare ``storm'' responsible for the cluster takes the form of an avalanche, where one event triggers another, which triggers another, etc., then the spreading out of the avalanche might not be perfectly axisymmetric.

Figure \ref{fig_nanoflares} shows a conceptual illustration of the described scenario by showing side-by-side images of a modified Model 2 with $\alpha = 1.7$ and $\gamma = 0.8$ and a synthetic EUV image produced by the above-mentioned MHD simulation. The parameters of the modified SPS model (see the column "Model 2a" in Table \ref{tab:stats}) are roughly consistent with those obtained for the Hi-C loops. It is worth noting  that even though our SPS models do not directly address the clustering effects present in the nanoflare simulation, their larger structures can be interpreted as bundles of smaller light-emitting structures associated with spatially-correlated nanoflare events. A more direct representation of spatial clustering will be included in the next implementation of the SPS framework. Also, for a more accurate reconstruction of a specific loop geometry, the SPS parameters can be fine-tuned to match the observational data using a rigorous optimization scheme. This methodology will be tested in a follow-up paper.

Irrespective of these upcoming improvements, the results reported here cast doubt on the possibility that a majority of the observed coronal loops are projections of large veils or MHD turbulence. Furthermore, if clusters of nanoflares do exist, our numerical tests unambiguously indicate that their envelopes cannot be highly non-circular. This being said, the possibility of small (perhaps sub-resolution) sheets is still open, as is the possibility of very large two-dimensional structures, such as structures of the size of the entire loop system, not addressed by our study. However, 
the intermediate range of scales associated with the majority of the observed loop systems is most likely dominated by true, and not apparent, quasi-one dimensional light-emitting structures. 

To answer the question posed in the title, our results suggest that most EUV coronal loops could be influenced, but definitely {\it not caused}, by three-dimensional projection effects. These effects can be isolated by properly chosen data analysis techniques, and what remains as the most likely underlying emitting geometry is a collection of quasi-one-dimensional luminous structures which we intuitively infer when observing coronal loops.

\hspace{2pt}

We thank the members of the Coronal Heating Team at NASA GSFC for stimulating discussions, N. Arge for helpful comments on the SPS method, A. Pouquet, P. Mininni and D. Rosenberg for the high-resolution turbulence data.  We acknowledge the High-resolution Coronal Imager (Hi-C 2.1) instrument team for making the second re-flight data available. This work was supported by the GSFC Heliophysics Internal Scientist Funding Model competitive work package program. VMU was also partly supported through the Partnership for Heliophysics and Space Environment Research (NASA grant No. 80NSSC21M0180).


\end{document}